\documentclass[twocolumn]{aastex62}

\usepackage{amsmath}
\usepackage{color}
\usepackage{xspace}

\newcommand{\hii}{H{\sc\,ii}\xspace}
\newcommand{\hi}{H{\sc\,i}\xspace}
\newcommand{\hm}{H$_2$\xspace}

\def \gastwo {\textsc{gasoline2}}

\def \apjl {ApJL}

\def \apjs {ApJS}
\def \aap {A\&A}

\def \spose#1{\hbox  to 0pt{#1\hss}}  
\def \lta{\mathrel{\spose{\lower 3pt\hbox{$\sim$}}\raise  2.0pt\hbox{$<$}}}
\def \gta{\mathrel{\spose{\lower  3pt\hbox{$\sim$}}\raise 2.0pt\hbox{$>$}}}
\def \ion#1#2{#1{\footnotesize{#2}}\relax}

\def \kms {\ifmmode  \,\rm km\,s^{-1} \else $\,\rm km\,s^{-1}  $ \fi }
\def \cm {\ifmmode  {\rm cm}  \else ${\rm  cm}$ \fi  }  
\def \kpc {\ifmmode  {\rm kpc}  \else ${\rm  kpc}$ \fi  }  
\def \hkpc {\ifmmode  {h^{-1}\rm kpc}  \else ${h^{-1}\rm kpc}$ \fi  }  
\def \hMpc {\ifmmode  {h^{-1}\rm Mpc}  \else ${h^{-1}\rm Mpc}$ \fi  }  
\def \Msun {\ifmmode \rm M_{\odot} \else $\rm M_{\odot}$ \fi}
\def \hMsun {\ifmmode h^{-1}\,\rm M_{\odot} \else $h^{-1}\,\rm M_{\odot}$ \fi}
\def \hhMsun {\ifmmode h^{-2}\,\rm M_{\odot}\else $h^{-2}\,\rm M_{\odot}$ \fi}
\def \Lsun {\ifmmode L_{\odot} \else $L_{\odot}$ \fi} 
\def \hhLsun {\ifmmode h^{-2}\,\rm L_{\odot} \else $h^{-2}\,\rm L_{\odot}$ \fi}

\def\LCDM{$\Lambda$CDM }
\def \LCDM {\ifmmode \Lambda{\rm CDM} \else $\Lambda{\rm CDM}$ \fi}
\def \sig8 {\ifmmode \sigma_8 \else $\sigma_8$ \fi} 
\def \Omegam {\ifmmode \Omega_{\rm m} \else $\Omega_{\rm m}$ \fi} 
\def \Omegab {\ifmmode \Omega_{\rm b} \else $\Omega_{\rm b}$ \fi} 
\def \Omegar {\ifmmode \Omega_{\rm r} \else $\Omega_{\rm r}$ \fi} 
\def \fbar {\ifmmode f_{\rm b} \else $f_{\rm b}$ \fi} 
\def \OmegaL {\ifmmode \Omega_{\rm \Lambda} \else $\Omega_{\rm \Lambda}$\fi} 
\def \Deltavir {\ifmmode \Delta_{\rm vir} \else $\Delta_{\rm vir}$ \fi}
\def \rhocrit {\ifmmode \rho_{\rm crit} \else $\rho_{\rm crit}$ \fi}

\def \rs {\ifmmode r_{\rm s} \else $r_{\rm s}$ \fi} 
\def \rh {\ifmmode r_{\rm h} \else $r_{\rm h}$ \fi} 
\def \Rvir {\ifmmode R_{\rm vir} \else $R_{\rm vir}$ \fi}
\def \Vvir {\ifmmode V_{\rm  vir} \else  $V_{\rm vir}$  \fi} 
\def \Vmax {\ifmmode V_{\rm  max} \else  $V_{\rm max}$  \fi} 
\def \Mvir {\ifmmode M_{\rm  vir} \else $M_{\rm  vir}$ \fi}  
\def \Mhalo {\ifmmode M_{200} \else $M_{200}$ \fi}  
\def \Nvir {\ifmmode N_{\rm  vir} \else $N_{\rm  vir}$ \fi}  
\def \Jvir {\ifmmode J_{\rm vir} \else $J_{\rm vir}$ \fi} 
\def \Evir {\ifmmode E_{\rm vir} \else $E_{\rm vir}$ \fi} 
\def \lam {\ifmmode \lambda  \else $\lambda$ \fi} 
\def \lamp {\ifmmode \lambda^{\prime} \else $\lambda^{\prime}$  \fi} 
\def \lampc {\ifmmode \lambda^{\prime}_{\rm c} \else
  $\lambda^{\prime}_{\rm c}$  \fi} 

\def \xoff {\ifmmode x_{\rm off} \else $x_{\rm off}$ \fi}
\def \rhorms {\ifmmode \rho_{\rm rms} \else $\rho_{\rm rms}$ \fi}
\def \qbar {\ifmmode \bar{q} \else $\bar{q}$ \fi}

\def \Mb {\ifmmode M_{\rm b} \else $M_{\rm b}$ \fi} 
\def \eSF {\ifmmode \epsilon_{\rm SF} \else $\epsilon_{\rm SF}$ \fi} 
\def \Md {\ifmmode M_{\rm d} \else $M_{\rm d}$ \fi} 
\def \Mg {\ifmmode M_{\rm g} \else $M_{\rm g}$ \fi} 
\def \Rb {\ifmmode R_{\rm b} \else $R_{\rm b}$ \fi} 
\def \Rd {\ifmmode R_{\rm d} \else $R_{\rm d}$ \fi} 
\def \Rg {\ifmmode R_{\rm g} \else $R_{\rm g}$ \fi} 
\def \mgal {\ifmmode m_{\rm gal} \else $m_{\rm gal}$ \fi} 
\def \rj {\ifmmode {\cal R}_j \else ${\cal R}_j$ \fi} 
\def \lamgal {\ifmmode \lambda_{\rm gal} \else $\lambda_{\rm gal}$ \fi} 
\def \Vcirc {\ifmmode V_{\rm circ} \else $V_{\rm circ}$ \fi} 
\def \Vrot {\ifmmode V_{\rm rot} \else $V_{\rm rot}$ \fi} 
\def \Vflat {\ifmmode V_{\rm flat} \else $V_{\rm flat}$ \fi} 
\def \Mstar {\ifmmode M_{\rm star} \else $M_{\rm star}$ \fi} 
\def \Mgas {\ifmmode M_{\rm gas} \else $M_{\rm gas}$ \fi} 
\def \Mbar {\ifmmode M_{\rm bar} \else $M_{\rm bar}$ \fi}
\def \Rbar {\ifmmode R_{\rm bar} \else $R_{\rm bar}$ \fi} 

\def \DeltaIMF {\ifmmode \Delta_{\rm IMF} \else $\Delta_{\rm IMF}$ \fi}

\def \VV {\ifmmode V_{\rm 2.2}/V_{200} \else $V_{2.2}/V_{200}$ \fi} 
\def \dvr {\ifmmode \partial_{\rm VR} \else $\partial_{\rm VR}$ \fi}

\received{\today}
\revised{\today}
\accepted{\today}
\submitjournal{ApJ}

\shorttitle{Gas fraction - q relation}
\shortauthors{Wang et al.}

\begin{document}

\title{Dynamic equilibrium sets atomic content of galaxies across cosmic time}

\correspondingauthor{Liang Wang}
\email{liang.wang@uwa.edu.au}

\author{Liang Wang}
\affil{International Centre for Radio Astronomy Research (ICRAR), M468, University of Western Australia, 35 Stirling Hwy, Crawley, WA 6009, Australia}

\author{Danail Obreschkow}
\affil{International Centre for Radio Astronomy Research (ICRAR), M468, University of Western Australia, 35 Stirling Hwy, Crawley, WA 6009, Australia}

\author{Claudia D.P. Lagos}
\affil{International Centre for Radio Astronomy Research (ICRAR), M468, University of Western Australia, 35 Stirling Hwy, Crawley, WA 6009, Australia}
\affil{ARC Centre of Excellence for All Sky Astrophysics in 3 Dimensions (ASTRO 3D)}

\author{Sarah M. Sweet}
\affil{Centre for Astrophysics and Supercomputing, Swinburne University of Technology, P.O.Box 218, Hawthorn, VIC 3122, Australia}

\author{Deanne B. Fisher}
\affil{Centre for Astrophysics and Supercomputing, Swinburne University of Technology, P.O.Box 218, Hawthorn, VIC 3122, Australia}

\author{Karl Glazebrook}
\affil{Centre for Astrophysics and Supercomputing, Swinburne University of Technology, P.O.Box 218, Hawthorn, VIC 3122, Australia}

\author{Andrea V. Macci{\`o}}
\affil{New York University Abu Dhabi, P.O.Box 129188, Saadiyat Island, Abu Dhabi, United Arab Emirates}
\affil{Max-Planck-Institut f\"ur Astronomie, K\"onigstuhl 17, D-69117 Heidelberg, Germany}

\author{Aaron A. Dutton}
\affil{New York Univsersity Abu Dhabi, P.O.Box 129188, Saadiyat Island, Abu Dhabi, United Arab Emirates}

\author{Xi Kang}
\affil{Purple Mountain Observatory, the Partner Group of MPI f\"ur Astronomie, 2 West Beijing Road, Nanjing 210008, China}



\begin{abstract}

We analyze 88 independent high-resolution cosmological zoom-in simulations of disk galaxies in the NIHAO simulations suite to explore the connection between the atomic gas fraction and angular momentum of baryons throughout cosmic time. The study is motivated by the analytical model of \citet{obreschkow16}, which predicts a relation between the atomic gas fraction $f_{\rm atm}$ and the integrated atomic stability parameter $q \equiv j\sigma / (GM)$, where $M$ and $j$ are the mass and specific angular momentum of the galaxy (stars+cold gas) and $\sigma$ is the velocity dispersion of the atomic gas. We show that the simulated galaxies follow this relation from their formation ($z\simeq4$) to present within $\sim 0.5$ dex. To explain this behavior, we explore the evolution of the local Toomre stability and find that $90\%$--$100\%$ of the atomic gas in all simulated galaxies is stable at any time. In other words, throughout the entire epoch of peak star formation until today, the timescale for accretion is longer than the timescale to reach equilibrium, thus resulting in a quasi-static equilibrium of atomic gas at any time. Hence, the evolution of $f_{\rm atm}$ depends on the complex hierarchical growth history primarily via the evolution of $q$. An exception are galaxies subject to strong environmental effects.

\end{abstract}




\section{Introduction} \label{sec:intro}

A comprehensive theory of galaxy evolution requires understanding the assembly and evolution of the stellar disks and spheroids of galaxies, as well as the co-evolution of these components with the interstellar medium (ISM) and circumgalactic medium (CGM). The accurate modeling of these gaseous components in galaxies is challenging as the gas is subject to non-linear gravitational, hydrodynamic and radiative forces. Several physical processes significantly affect the geometry and thermodynamic phase of the gas, such as cold flow accretion \citep{keres05}, hot mode accretion \citep[e.g.][]{rees77, white78, putman12, werk14}, stellar winds from evolved stars \citep{kalirai08, leitner11} and recycling of the metal-rich gas ejected through stellar winds \citep{oppenheimer10, brook14, ubler14}. Owing to the time-dependent complex geometry of gas flows into and out of galaxies, the detailed evolution of different gas components has yet to be understood.

Neutral atomic hydrogen (\hi) dominates the hydrogen budget in local galaxies, except at the highest column densities ($>10 ~\Msun~pc^{-2}$), where this gas normally transitions into the molecular (\hm) phase. \hi is the critical waypoint between the ionized CGM and star formation in the disk \citep{leroy08}. Detailed studies of \hi are therefore invaluable to understanding the formation of galaxies at large. Direct observations in 21cm emission and absorption \citep{ewen51} have revealed a plethora of relationships between the \hi content and other galaxy properties, most notably the star-formation rate \citep{kennicutt89}, stellar mass \citep[e.g.][]{dutton09, obreschkow09, dutton11, catinella13,maddox15}, spin \citep{huang12,obreschkow16} and morphology \citep{catinella10, brown15, brown17}.



The atomic gas fraction is defined as:
\begin{equation}\label{eq:fatm_def}
	f_{\rm atm} = \frac{1.35M_{\rm \ion{H}{I}}}{M},
\end{equation}
where the total mass $M=M_\star+1.35(M_{\rm HI}+M_{\rm H_2})$, $M_\star$, $M_{\rm HI}$ and $M_{\rm H_2}$ are stellar mass, \hi mass and \hm mass respectively. The factor of $1.35$ accounts for the universal $\sim 26\%$ helium fraction at redshift $z=0$. Computational examinations show that $f_{\rm atm}$ depends sensitively on the numerical resolution, subgrid physics, e.g. feedback from supernovae and active galactic nuclei, \citep[e.g.][]{duffy12, dave13, stinson15, crain17, diemer18} and physical processes related to the cosmological environment, e.g. ram pressure stripping and tidal interactions, \citep{cunnama14,  rafieferantsoa15}. It is necessary to identify the key driver(s) that set(s) $f_{\rm atm}$ to first order in some well-defined sense.

Several recent empirical and computational works have highlighted that the specific angular momentum of galaxies at fixed stellar mass is strongly correlated to their atomic gas fraction \citep[e.g.][]{dutton10, huang12,obreschkow15b,lagos17, romeo18, stevens18, zoldan18}. 

\citet{obreschkow16} (hereafter O16) introduced a parameter-free analytical model that predicts $f_{\rm atm}$ as a function of mass and angular momentum in equilibrium disks. This model assumes that galactic disks have an exponential surface density profile and are locally either fully atomic or non-atomic: the disk is atomic where and only where the atomic gas is stable in the sense of \citet{toomre64} at the characteristic dispersion velocity $\sigma$ of the warm neutral medium (about $10~\rm km\,s^{-1}$). In this model $f_{\rm atm}$ only depends on the so-called \emph{integrated atomic stability parameter}
\begin{equation}\label{eq:q}
	q = \frac{j_{\rm gal}\sigma_{\rm gas}}{GM_{\rm gal}},
\end{equation}
first introduced by \citet{obreschkow14}, where $M_{\rm gal}$ and $j_{\rm gal}$ are the mass and specific angular momentum (AM) of the galaxy (stellar disk+cold gas) and $G$ is the gravitational constant. O16 predict that $f_{\rm atm}$ depends on $q$, approximately as
\begin{equation}\label{eq:fatm}
	f_{\rm atm} = {\rm min}\{1, 2.5q^{1.12}\}
\end{equation}
with small ($<10\%$) variations subject to the shape of the rotation curve.

To the extent that the assumptions of O16 remain valid across cosmic time, the evolution of $f_{\rm atm}$ should depend on a galaxy's complex assembly and interaction history only (or at least predominantly) via the evolution of $q$. This hypothesis is an interesting test case for cosmological simulations, which provide comprehensive information on the history of the atomic gas in evolving galaxies. The aim of this study is to examine the dependency between $f_{\rm atm}$ and $q$ across the cosmic time in the Numerical Investigation of a Hundred Astrophysical Objects, NIHAO \citep{wang15} project. The NIHAO simulations are a suite of 88 hydrodynamical cosmological zoom-in simulations implementing the tree-smoothed particle hydrodynamics (SPH), \gastwo.  The NIHAO runs keep the same stellar physics at the whole mass range. The stellar mass of each halo in the NIHAO sample agrees with the prediction from abundance matching \citep{wang15}. The galaxies in the NIHAO sample reproduce several baryonic properties in observation, such as the star formation main sequence \citep{wang15}, the column density profile of cool \hi \citep{gutcke17}, the Tully-Fisher relation \citep{dutton17} and the local velocity function \citep{maccio16}. Therefore, NIHAO is well suited to study the relation (if any) between $f_{\rm atm}$ and $q$ through cosmic time across six orders of magnitude in stellar mass from $10^5 \Msun$ to $10^{11} ~\Msun$.

This paper is structured as follows. The simulation techniques, in particular the modelling of the different hydrogen phases and computation of relevant kinematic parameters, are described in Section~\ref{sec:sim}. The properties of the simulated galaxies and the key results concerning the relation between the atomic gas fraction and $q$ parameter are presented in Section~\ref{sec:revisit}, along with a discussion of the physical mechanisms leading to these results.
A summary and outlook are given in Section~\ref{sec:summary}.

\section{Simulations} \label{sec:sim}

This section gives an overview of the NIHAO simulations and briefly describes the subgrid physics routines, including the scheme used to separate the hydrogen into ionized (\hii), atomic and molecular phases. We also describe the methods used to compute the kinematic parameters used in our analysis.

\subsection{Simulations and subgrid physics}

In this study, we use 88 zoom-simulations from the NIHAO project \citep{wang15}. In these simulations, the particle mass of the cold dark matter (CDM) and gas particles depends on the galaxy mass, such that each galactic halo is resolved by roughly $10^6$ CDM particles at redshift $z=0$. These zoom volumes have been extracted from three different $N$-body CDM simulations with a box size of 60, 20 and 15 \hMpc, respectively (see \citealt{dutton14} for details). All these runs used the cosmological parameters from the {\it Planck} satellite \citep{planck14}. Dark matter particle masses range from $\sim 10^4 ~\Msun$ in our lowest mass halos to $\sim 10^6 ~\Msun$ in our most massive halos, and their force softening lengths range from $\sim 150$ to $\sim 900$ pc, respectively. Gas particles are less massive by a factor of $(\Omega_{\rm dm} / \Omega_{\rm b}) \simeq 5.48$, where $\Omega_{\rm dm}$ and $\Omega_{\rm b}$ are density parameters of dark matter and baryon, and the corresponding force softening lengths are 2.34-times smaller. 

The simulated galaxies uniformly cover a range in stellar mass of $10^5 \lesssim M_\star / \Msun \lesssim 10^{11}$ at $z=0$. Most systems of $M_\star < 10^9 ~\Msun$ are rotationally dwarfs with disky stellar and cold gas morphology, sometimes showing typical irregularities of dwarfs. Most of the more massive galaxies are spiral systems with rotating central over-densities (pseudo-bulges) (Wang et al. submitted). A few most massive galaxies are early-type systems dominated by a spheroid. Two systems have undergone a major merger just before $z=0$, and show significant merger remnant structures.

We use the smoothed particle hydrodynamics code \gastwo ~\citep{wadsley17}. The code includes a subgrid model for turbulent mixing of metal and energy \citep{wadsley08}, ultraviolet (UV) heating, photo-ionization and cooling due to hydrogen, helium and metals \citep{shen10}.

The star formation and feedback models are those used in the Making Galaxies in a Cosmological Context (MaGICC) simulations \citep{stinson13}. The gas is converted into stars according to the Kennicutt-Schmidt law when it satisfies a temperature and density threshold. Stars feed both metals and energy back in to the ISM gas surrounding the region where they formed. Supernova (SN) feedback is implemented using the blastwave formalism described in \citet{stinson06}. Pre-SN feedback is an attempt to consider radiation energy from massive stars. Heating is introduced immediately after massive stars form based on how much star light is radiated. Our simulations use thermal feedback to provide pressure support and increase gas temperature above the star formation threshold, and thus to decrease star formation. Full details on the star formation and feedback modeling can be found in \citet{wang15}.

\subsection{Partition of hydrogen into \hii, \hi and \hm}\label{ss:partition}

\begin{figure*}[ht!]
\epsscale{1.15}
\includegraphics[width=0.5\textwidth]{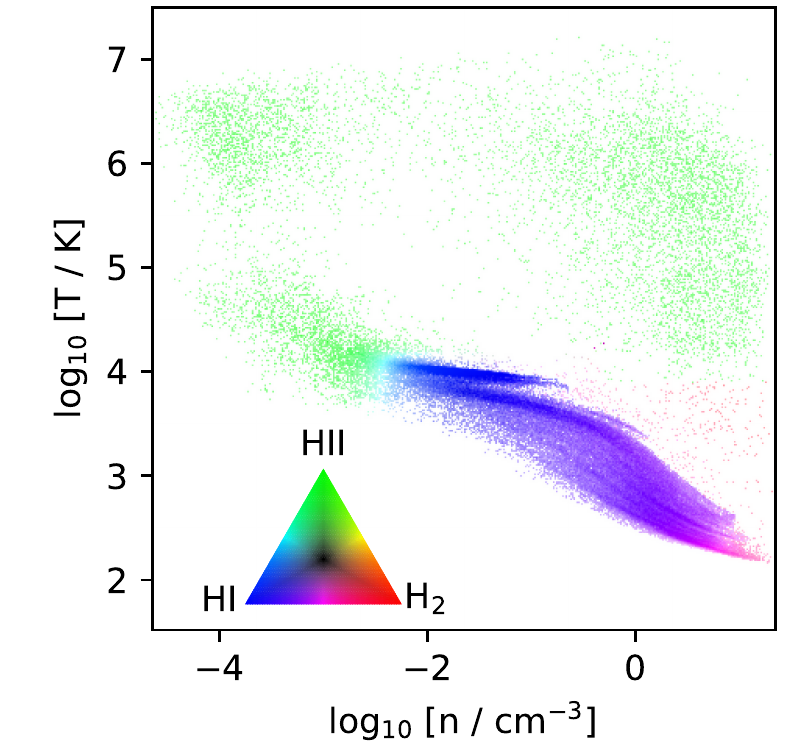}
\includegraphics[width=0.5\textwidth]{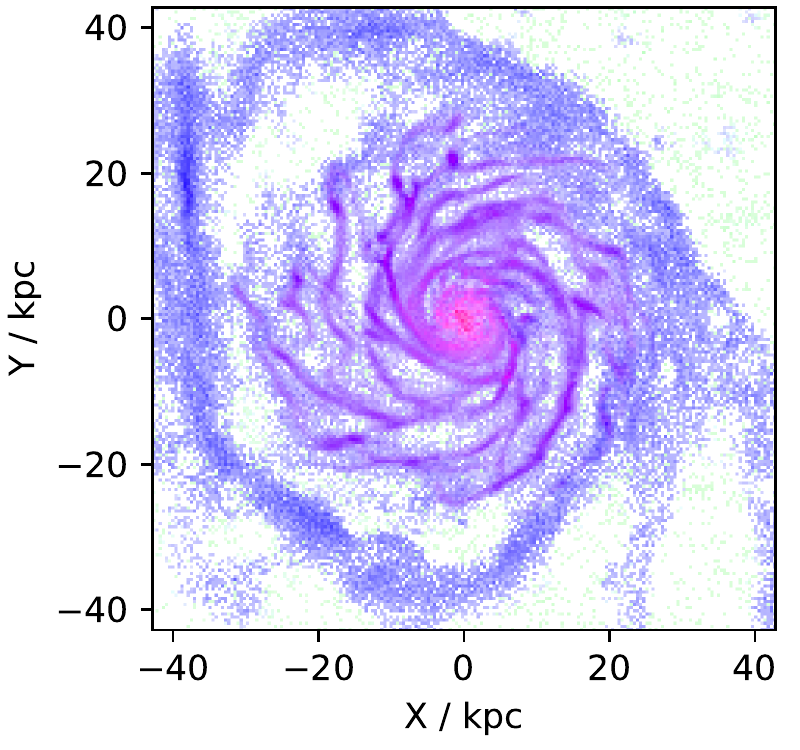}
\caption{Temperature-density phase-diagram (left) and spatial distribution (right) of all the hydrogen in the Milky Way-like NIHAO galaxy g8.26e11 at $z=0$. In each pixel, the balance between the three hydrogen phases is represented by hue according to the triangle, while intensity represents the total hydrogen mass on a non-linear scale ($\gamma=0.5$) to show low-density regions.
         \label{fig:phase_dia}}
\end{figure*}

The partitioning of the gas particles into \hii, \hi and \hm is done following a two-stage scheme, similar to those presented by \citet{rahmati15, lagos15, lagos16, bahe16,crain17}. Firstly, for the transition from \hii to neutral (\hi+\hm) gas, we use the fitting function of \citet{rahmati13} to calculate the neutral fraction on a particle-by-particle basis from the gas temperature, gas density and the UV background modeled by \citet{haardt01}. This fitting function accounts for collisional ionization, photo-ionization by a homogeneous UV background and radiative recombination. Secondly, the neutral gas particles are fractionally divided into \hi and \hm using the model of \citet{gnedin11}. This model relies on a phenomenological model for \hm formation, approximating how \hm forms on the surfaces of dust grains and is destroyed by the interstellar radiation field. In this model, the \hm / \hi ratio of individual gas particles depends on the dust-to-gas mass ratio, gas surface density and UV field, which we calculated as in \citet{lagos15}. \citet{lagos15} used the models of \citet{krumholz13} and \citet{gnedin14} to calculate the \hm fraction for individual particles, finding similar results. \citet{diemer18} models the UV radiation from young stars by assuming a constant escape fraction and optically thin propagation throughout the galaxy and improves the calculation of \hm mass. Our test cases show that the partitioning scheme in \citet{diemer18} provides similar \hi mass as well.

The phase partitioning of hydrogen is illustrated in Fig.~\ref{fig:phase_dia} for a Milky Way-like galaxy (NIHAO object g8.26e11) at $z=0$, both in the temperature-density phase-space, as well as in the real space phase-on projection of the galaxy. In each pixel of the two panels, hue represents the phase mixing and intensity represents the total hydrogen density. As expected, most \hm is found in the dense center and spiral arms of the galaxy, whereas the \hi dominates in the outskirts.

The majority of the hydrogen in this galaxy resides in the high-density ($n > 10^{-3} {\rm ~cm}^{-3}$) and low-temperature ($T < 10^4$ K) region of the phase-diagram, where the material is almost exclusively neutral. This statement only applies to the region of the galaxy and does not conflict with the likely fact that most of the hydrogen in the universe resides in the ionized circum-galactic medium (CGM) or intergalactic medium (IGM) \citep{crain17}.

Because of the limited resolution, the interstellar medium gas particles around supernovae are always dense and would quickly radiate their energy away due to efficient cooling at high density. For this reason, cooling is disabled for particles inside the blast region (for a duration specified in \citealp{mckee77}). The locally disabled cooling artificially maintains too much ionized hydrogen in the high-density ($n>1~{\rm cm}^{-3}$), high-temperature ($T > 10^4$ K) state. We consider this gas to be always ionised, and thus this is of no concern for the present analysis of atomic hydrogen.


\begin{figure*}[ht!]
\epsscale{1.15}
\includegraphics[width=0.5\textwidth]{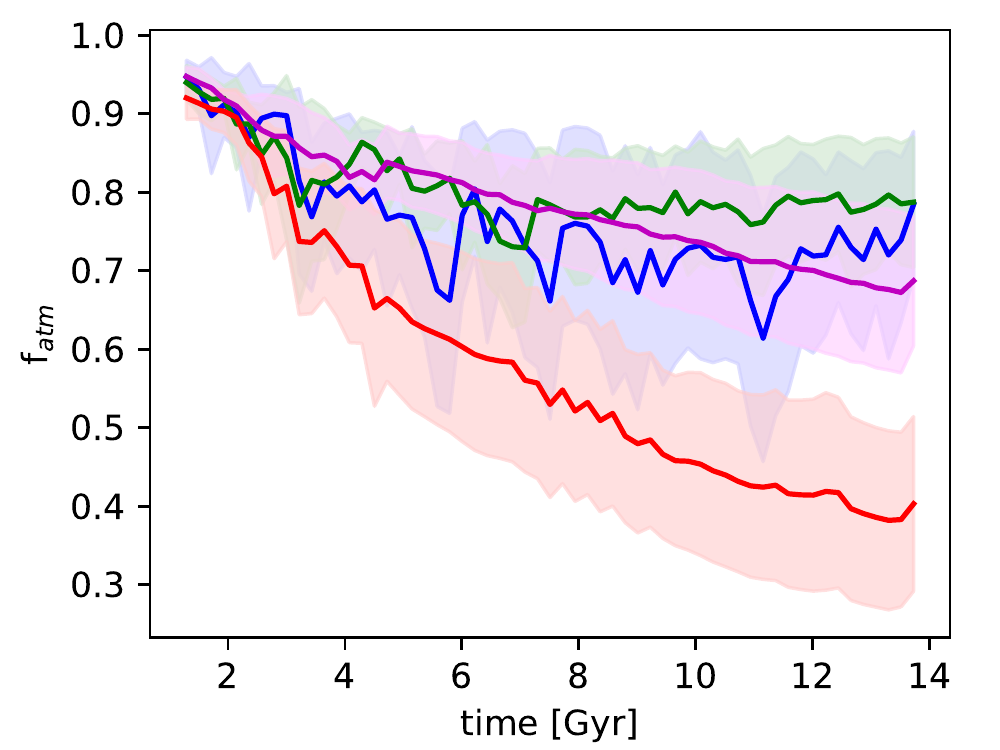}
\includegraphics[width=0.5\textwidth]{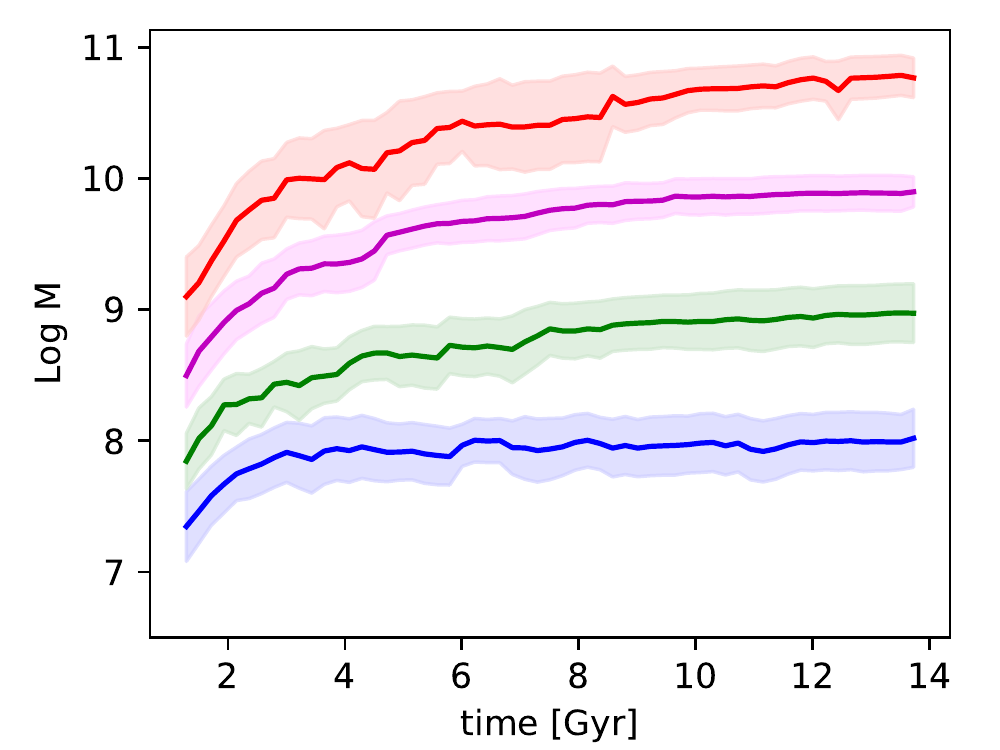}
\includegraphics[width=0.5\textwidth]{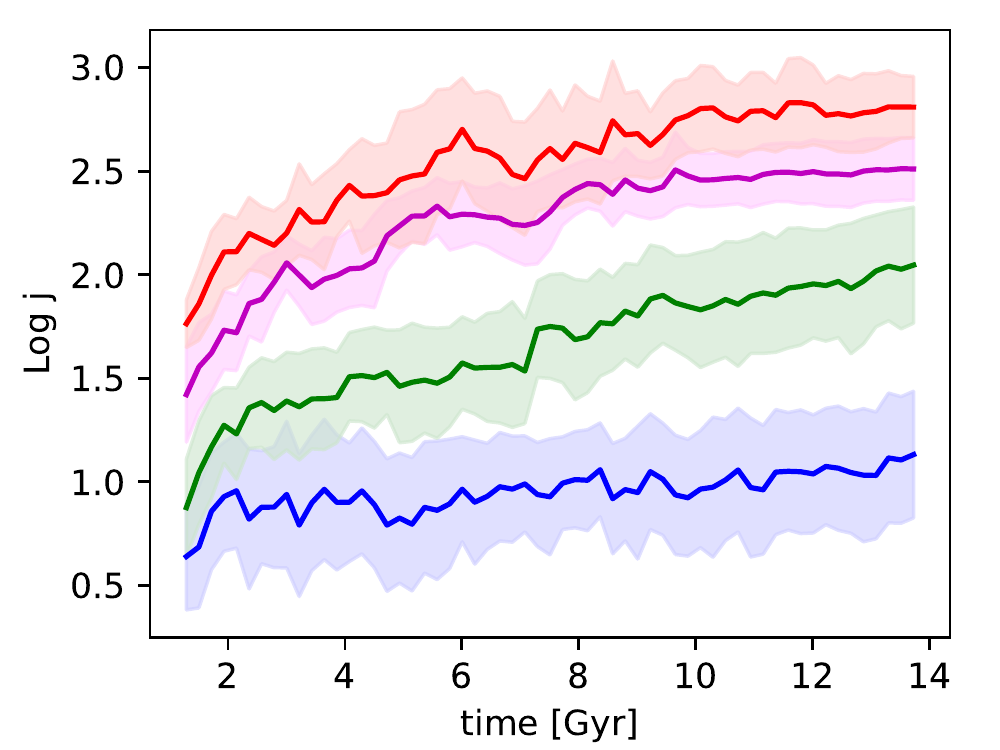}
\includegraphics[width=0.5\textwidth]{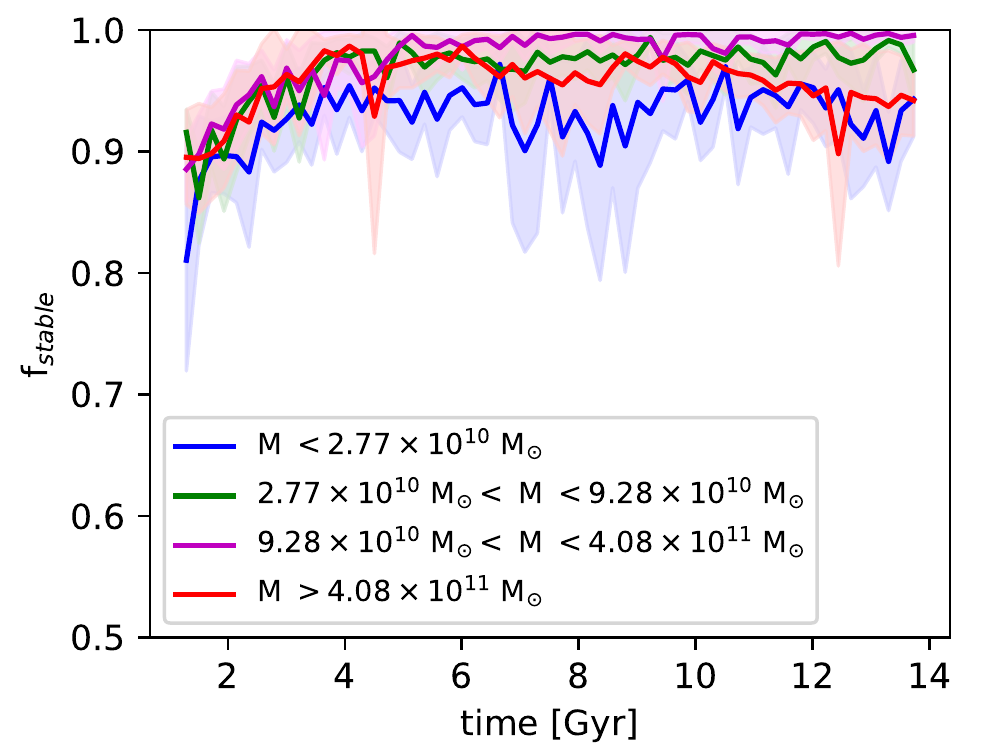}
\caption{Average evolution of atomic gas fractions (upper left), baryon mass (upper right), specific angular momenta (lower left), and stable mass fraction (lower right) of the 88 NIHAO galaxies in four bins of virial mass at $z=0$. The shaded regions show the 1$\sigma$ standard deviations. The three lower mass bins ($<4.08\times10^{11} ~\Msun$) show a similar evolution, with an atomic gas fraction of 70\%--80\% at $z=0$. The most massive galaxies in the sample, however, have an atomic gas fraction that decreases  steeply with time and become stellar mass-dominated at $t \gtrsim 8 $ Gyr. Baryonic mass and specific AM can vary strongly and systematically between different mass bins and generally increase with time. In all mass bins, the stable mass fraction lies above $\sim 90\%$ at any time shortly after the galaxies form.\vspace{4mm}}\label{fig:review}
\end{figure*}

\begin{figure}[b]
\epsscale{1.15}
\plotone{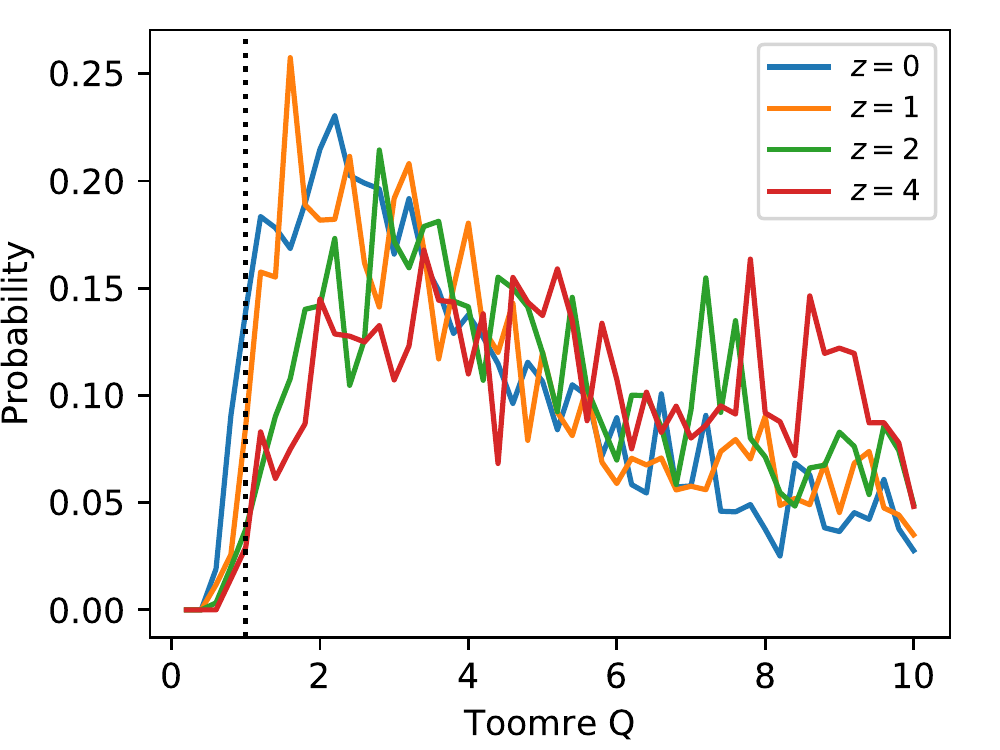}
\caption{Mean distribution of the net two-component Toomre $Q$ (Appendix A) for all simulated galaxies. The distribution function of $Q$ across the galaxies highlights that most of the stable gas lies significantly above $Q=1$ at all redshifts.
         \label{fig:tq_dis}}
\end{figure}

\subsection{The q Parameters}

This section describes the methods to compute the quantities needed to study the $q$--$f_{\rm atm}$ relation of O16. To study this relation, we must calculate the atomic gas velocity dispersion $\sigma$, galaxy mass and angular momentum. All these quantities are measured in the galactic region confined to a flat cylinder aligned with the galactic plane, of radius $5 R_{50}$ and height $0.2 R_{50}$, where $R_{50}$ is the stellar half-mass radius of that galaxy. In principle, O16 only account for a disk component, hence $M$ and $j$ in this study should perhaps exclude bulge stars, although in the case of disk-like pseudo-bulges the choice is not straightforward. In this paper, we do not decompose galaxies into disks and bulges and simply include all stellar material in $M$ and $j$. This represents at most a modest error, since our galaxies are disk-dominated or even bulge-less at lower stellar masses ($M_\star<10^{10}~\Msun$). Note that elliptical galaxies generally exhibit very low atomic gas fractions ($f_{\rm atm}<0.01$), negligible for the cosmic \hi budget.

The neutral atomic gas fraction $f_{\rm atm}$ is simply calculated via eq.~(\ref{eq:fatm_def}), where the  \hi and \hm masses result directly from the phase splitting of Section \ref{ss:partition}.

The specific angular momentum $j$ of the galaxy is computed as
\begin{equation}
j = \frac{|\sum M_i  \mathbf{r}_i \times \mathbf{v}_i|}{\sum M_i},
\end{equation}
where $i$ goes over all baryonic particles in the cylindrical region, $M_i$ are the particle masses (stellar+\hi+\hm, excluding \hii), $\mathbf{r}_i$ are the position vectors from the center of mass, and $\mathbf{v}_i$ are the velocities relative to the center of mass frame. Given $j$, we then evaluate $q$ via eq.~(\ref{eq:q}). Because the disk thickness can effect the stability \citep{romeo13}, it is reasonable to measure the 3 dimensional dispersion to take the thinkness into account. Therefore, Unlike in O16, we do not assume a universal value for the atomic dispersion $\sigma$, but instead compute this quantity across the galaxy disk from the simulation as described in appendix A.

Finally, we quantify the stable mass fraction of the atomic hydrogen. A thin disk in gravitational equilibrium is stable if and only if the so-called Toomre parameter $Q$ \citep{toomre64} is larger than unity. For a single-component gaseous disk, this parameter takes the form $Q_{\rm gas} = \kappa \sigma / \pi G \Sigma$, where $\sigma$ and $\Sigma$ are the local radial velocity dispersion and surface density of the gas, respectively, and $\kappa$ is the local epicyclic frequency. A two-component (stellar+gas) stability parameter can be approximated using the formalism of \citet{romeo11}. This computation of $Q$ and the prerequisite computations of $\sigma$ and $\kappa$ are detailed in appendix A. Each galaxy is sub-divided in 400 cells, that is in 20 angular bins and 20 radial bins with approximately equal numbers of gas particles. In each cell $i$, the Toomre $Q_i$ is evaluated and the stable atomic gas fraction is computed as
\begin{equation}\label{eq:stable}
	f_{\rm stable} = M_{\rm HI}^{-1} \sum_{i\in{Q_i>1}} M_{{\rm HI},i}.
\end{equation}
Following this definition, $f_{\rm stable}$ is bound between 0 (all \hi unstable) and 1 (all \hi stable).

\section{Results and Discussion} \label{sec:revisit}

This section describes the cosmic evolution of $f_{\rm atm}$ of our 88 simulated galaxies, in relation to the cosmic evolution of other dynamical and kinematic parameters. 

\begin{figure}[t]
\epsscale{1.15}
\plotone{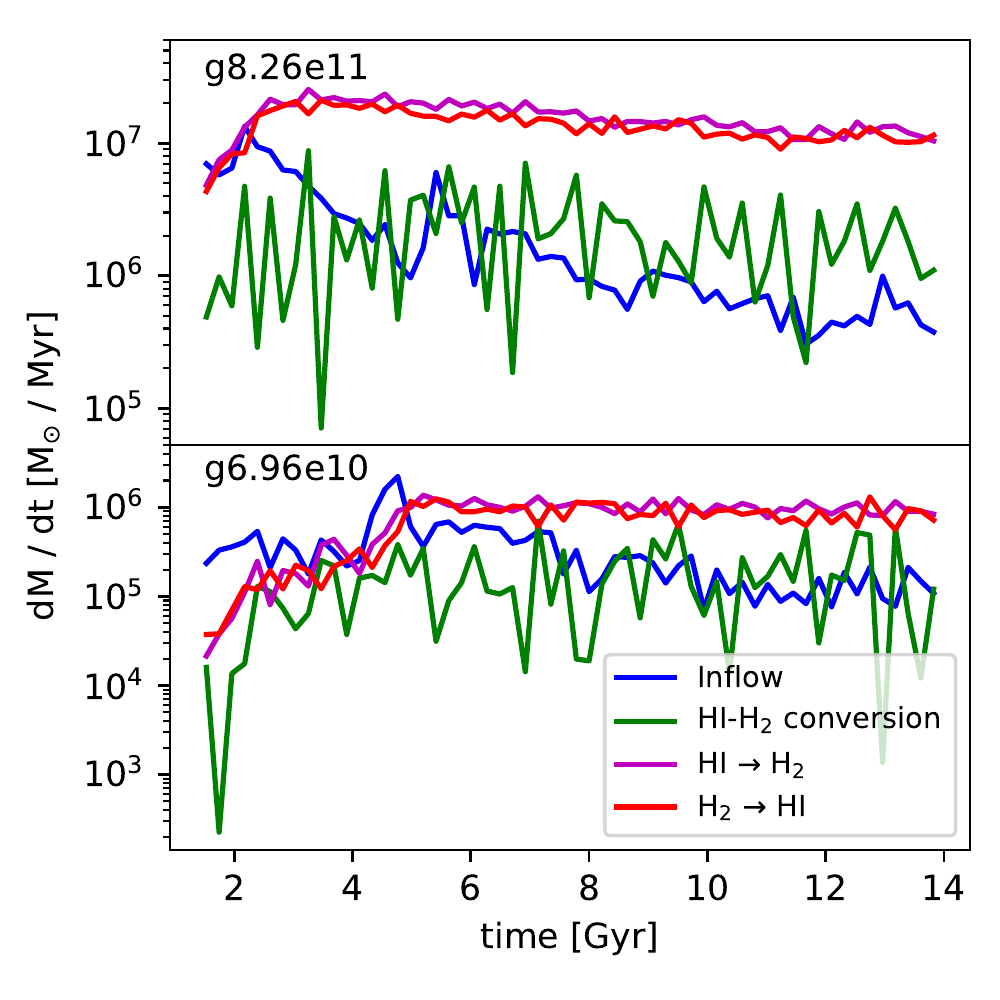}
\caption{Neutral hydrogen accretion rate (blue), \hi - \hm conversion rate (green) and mass flow rates from \hi to \hm (magenta) and viceversa (red) in two representative NIHAO galaxies (g8.26e11 in the Milky Way mass range and g6.96e10 in the dwarf galaxy range). The individual rates of the local molecularization (\hi$\rightarrow$~\hm) and feedback-driven dissociation (\hm$\rightarrow$\hi) are much larger than the resulting net \hi$\rightarrow$~\hm conversion rate. The \hi phase is in a quasi-static equilibrium at almost any time in the NIHAO galaxies. 
         \label{fig:m_flow}}
\end{figure}

\begin{figure*}[t]
\epsscale{1.15}
\plotone{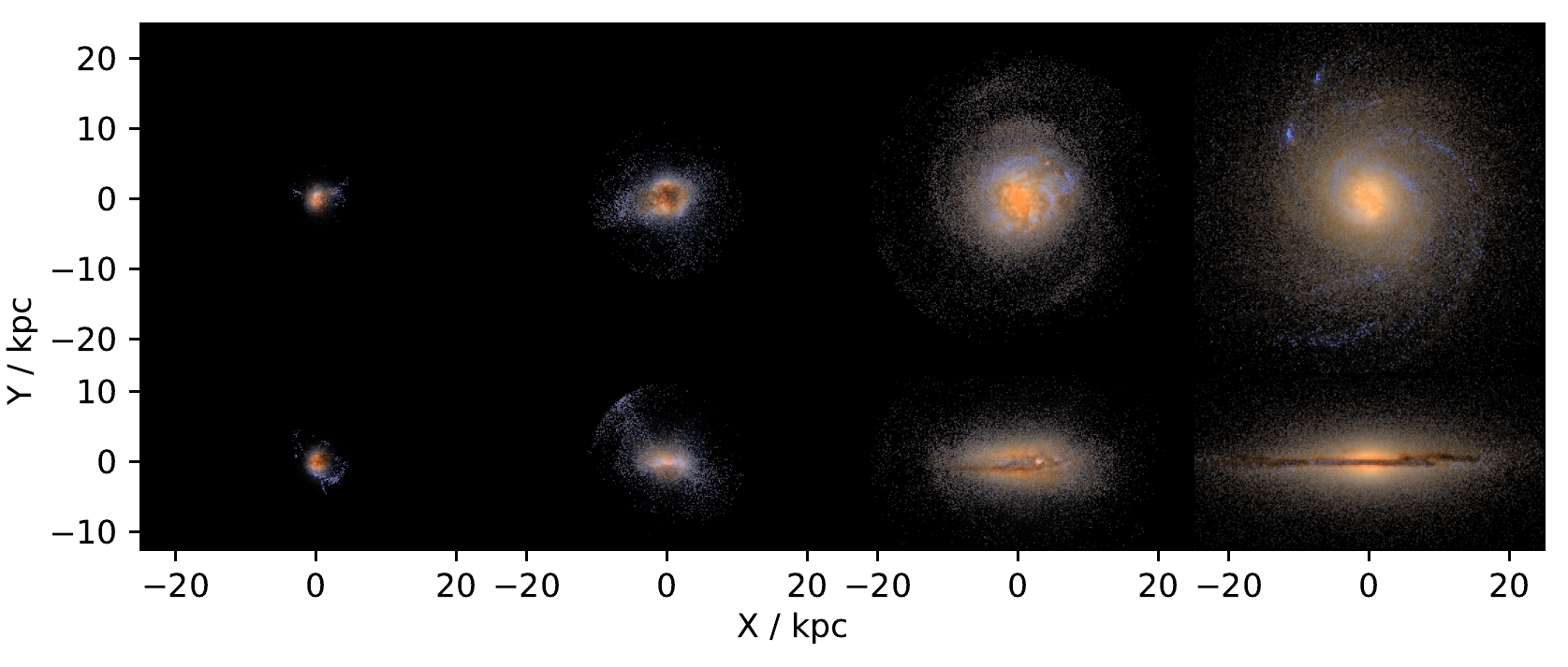}
\caption{Face-on (upper panel) and Edge-on (lower panel) views of g8.26e11 at redshift $z=4,2,1,0$ (left to right). The image is produced by post-processing through the Monte Carlo radiative transfer code SUNRISE. 
         \label{fig:image_evo}}
\end{figure*}

\subsection{Evolution of basic quantities} \label{sec:overview}

The upper left panel of Fig.~\ref{fig:review} shows the evolution of the mean atomic gas fraction in four bins of dynamical mass at $z=0$. We choose the bin breaks by equal numbers of galaxies. The general trend of the atomic gas fraction for all galaxies is to decrease with cosmic time, as expected due to the build-up of stellar mass. The three lower mass bins ($<4.08\times10^{11} ~\Msun$) show a similar evolution, with an atomic gas fraction of 70\%--80\% by $z=0$. The most massive galaxies in the sample, however, have an atomic gas fraction that decreases  steeply with time and become stellar mass-dominated at $t \gtrsim 8 $ Gyr. The co-evolution of the baryonic mass and specific AM are shown in the upper right and lower left panels of Fig.~\ref{fig:review}. Naturally, these quantities can vary strongly and systematically between different mass bins and generally increase with time. The evolutionary tracks of mass and AM are obviously distinct between different mass bins, while those of the atomic gas fraction $f_{\rm atm}$ are overlapping for the three lower mass bins. Hence, neither the mass nor the AM alone can determine the atomic gas fraction at any given cosmic time.

Following O16, we expect $f_{\rm atm}$ to correlate strongly with $q$. This expectation requires the atomic disks to be saturated in a stable equilibrium. We therefore explore the stable atomic gas fraction (eq.~\ref{eq:stable}) as a function of cosmic time in the lower right panel of Fig.~\ref{fig:review}. In all mass bins this fraction lies above $\sim 90\%$ at any time shortly after the galaxies form, when the universe was roughly $1~\rm Gyr$ old. Fig.~\ref{fig:tq_dis}, which shows the mean distribution function of $Q$ in all galaxies at different redshifts, highlights that most of the stable gas lies significantly above $Q=1$.

The reason for the high cold gas stability at all times is that the characteristic timescale of cold gas accretion onto the galaxies is almost always larger than that of the feedback-regulated \hi$\leftrightarrow$~\hm conversion loop in unstable regions. To illustrate this feature, we study the conversion rates in  MW and a dwarf galaxy. Fig.~\ref{fig:m_flow} shows three mass flow rates between different gas phases in two representative NIHAO galaxies (g8.26e11 in the Milky Way mass range and g6.96e10 in the dwarf galaxy range). 
The individual rates of the local molarization (\hi$\rightarrow$~\hm) and feedback-driven dissociation (\hm$\rightarrow$\hi) are much larger than the resulting net \hi$\rightarrow$~\hm conversion rate. The latter equals the cold gas accretion rate (\hii$\rightarrow$~\hi) onto the disk in the steady state situation, which roughly applies to our galaxies. In other words, the \hi phase is in a quasi-static equilibrium at almost any time in the NIHAO galaxies. 

This situation is also expected in real galaxies, as long as the time-scale of the local \hi$\rightarrow$~\hm conversion (i.e.~without accounting for feedback) is shorter than the time-scale of cold gas accretion onto the disk. We expect this to be the case in most spiral systems, where the local \hi$\rightarrow$~\hm conversion (before feedback) is similar to the local free-fall time order of $\sim 1 - 10$ Myr \citep{krumholz14}. However, this argument breaks down in very low-density and low-metallicity systems, where the \hm formation time can increase significantly \citep{krumholz13}, and hence the instantaneous self-regulation assumed in the O16 model breaks down. However, such galaxies are typically dwarf galaxies (high $q$ values), where this model predicts purely atomic disks anyways.

Incidentally, as discussed by \citet{stinson15}, the \hi mass in disky NIHAO galaxies remains approximately constant from $z=1$ to $z=0$. This means that at $z=1$ the quasi-static equilibrium reaches a state where the \hi accretion matches its depletion. In the present context this can be understood as a leveling-off in the evolution of $j/M$, which implies that newly accreted \hi will settle onto the existing \hi disk and hence reduce its stability until the same amount of \hi is converted into molecules.

In conclusion, the finding that most \hi is dynamically stable at any time, irrespective of the galaxy mass, motivates the analysis of $f_{\rm atm}$ in the framework of O16.




\subsection{Evolution of a single system in $q$-f$_{\rm atm}$ space} \label{sec:coevo}


Let us first consider the case of the single Milky Way-like galaxy (NIHAO object g8.26e11) already used in Fig.~\ref{fig:phase_dia}. We remind the reader that this galaxy is representative of the Milky-Way like galaxies in NIHAO, both in mass and morphology.  Fig.~\ref{fig:image_evo} shows the morphological evolution of this galaxy at four redshifts. The morphology at redshift $z=4$ is compact and irregular, due to turbulent initial collapse of low-angular momentum material. At $z=2$, this galaxy starts developing a disk, which becomes steadily more extended and dusty.

\begin{figure}[t]
\epsscale{1.15}
\plotone{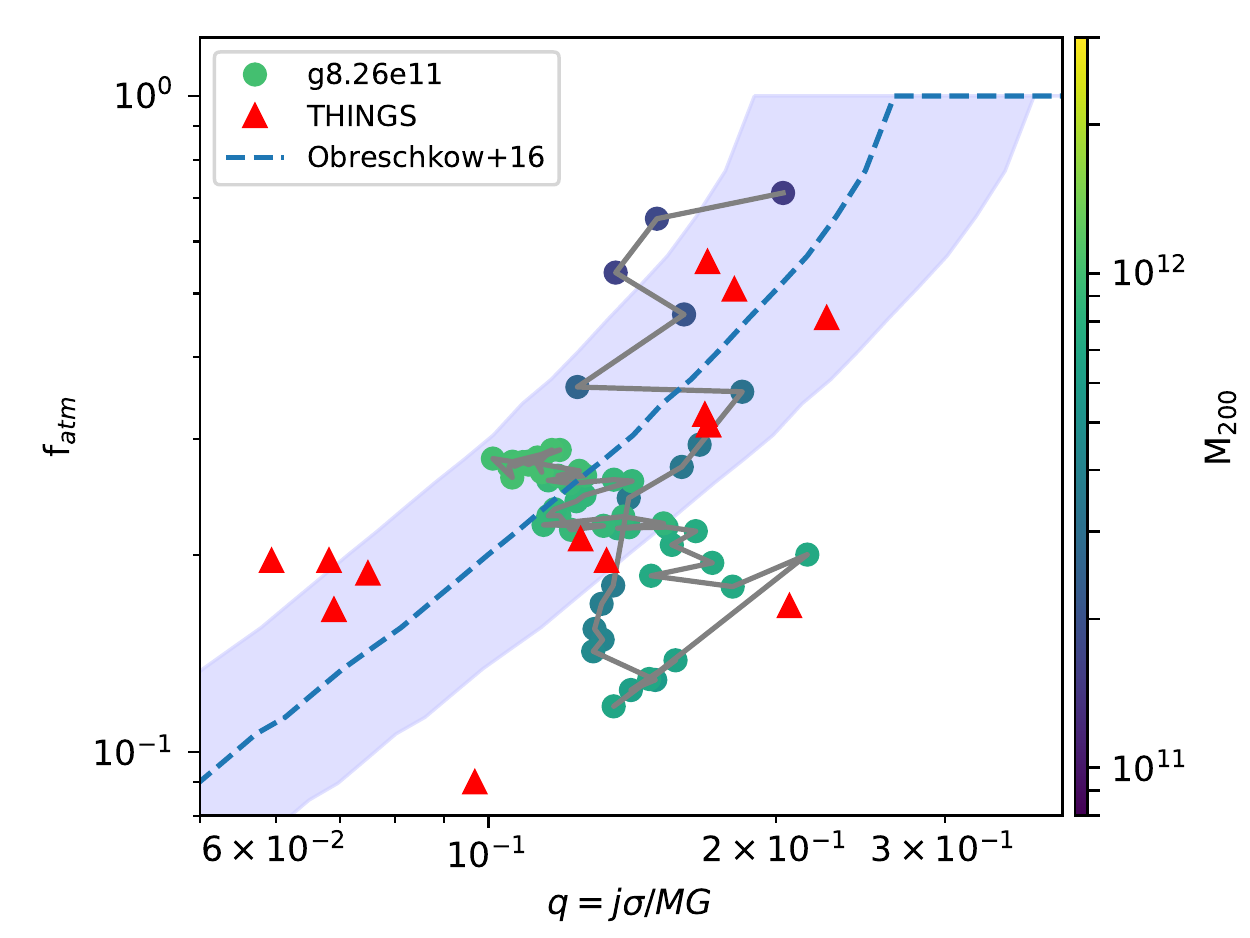}
\caption{Evolutionary track of the Milky Way-like NIHAO galaxy g8.26e11 in $q$-$f_{\rm atm}$ space from redshift $z=4$ to $z=0$ (connected round dots). The dots are color-coded by dynamical mass. Red triangles show the local spiral galaxies from the THINGS survey analyzed by O16 and the dashed line is eq.~\ref{eq:fatm} of their model.The evolutionary track of this galaxy in $q-f_{\rm atm}$ space appears to evolve around the relation of O16 roughly within its empirical scatter.}
         \label{fig:one_evo}
\end{figure}

The evolutionary track of this galaxy in $q-f_{\rm atm}$ space (see Fig.~\ref{fig:one_evo}) appears to scatter around the stability relation (dashed line). The galaxy starts somewhat above the relation, due to the fast accretion of \hi, not yet settled in a stable equilibrium disk. From there, $f_{\rm atm}$ first decreases dramatically due to disk heating by minor mergers (an effect discussed in detail by \citet{stevens18} in the context of semi-analytic models), making the galaxy \hi-deficient relative to the amount of \hi that could be dynamically supported. The galaxy evolves with low $f_{\rm atm}$ ($0.2$ dex lower than the predicted relation) for $\sim 1$ Gyr. Then the galaxy gradually accumulates new \hi, which mostly settles in a stable disk, moving this object steadily back onto the stability relation. The upper panel of Fig.~\ref{fig:m_flow} shows that the neutral gas accretion of this galaxy decreases monotonously before redshift $z \sim 2$, and keeps constant at the late stage. The decreasing efficiency of neutral gas accretion at early times couples with the decrease in $f_{\rm atm}$.


\begin{figure*}[ht]
\epsscale{1.15}
\plotone{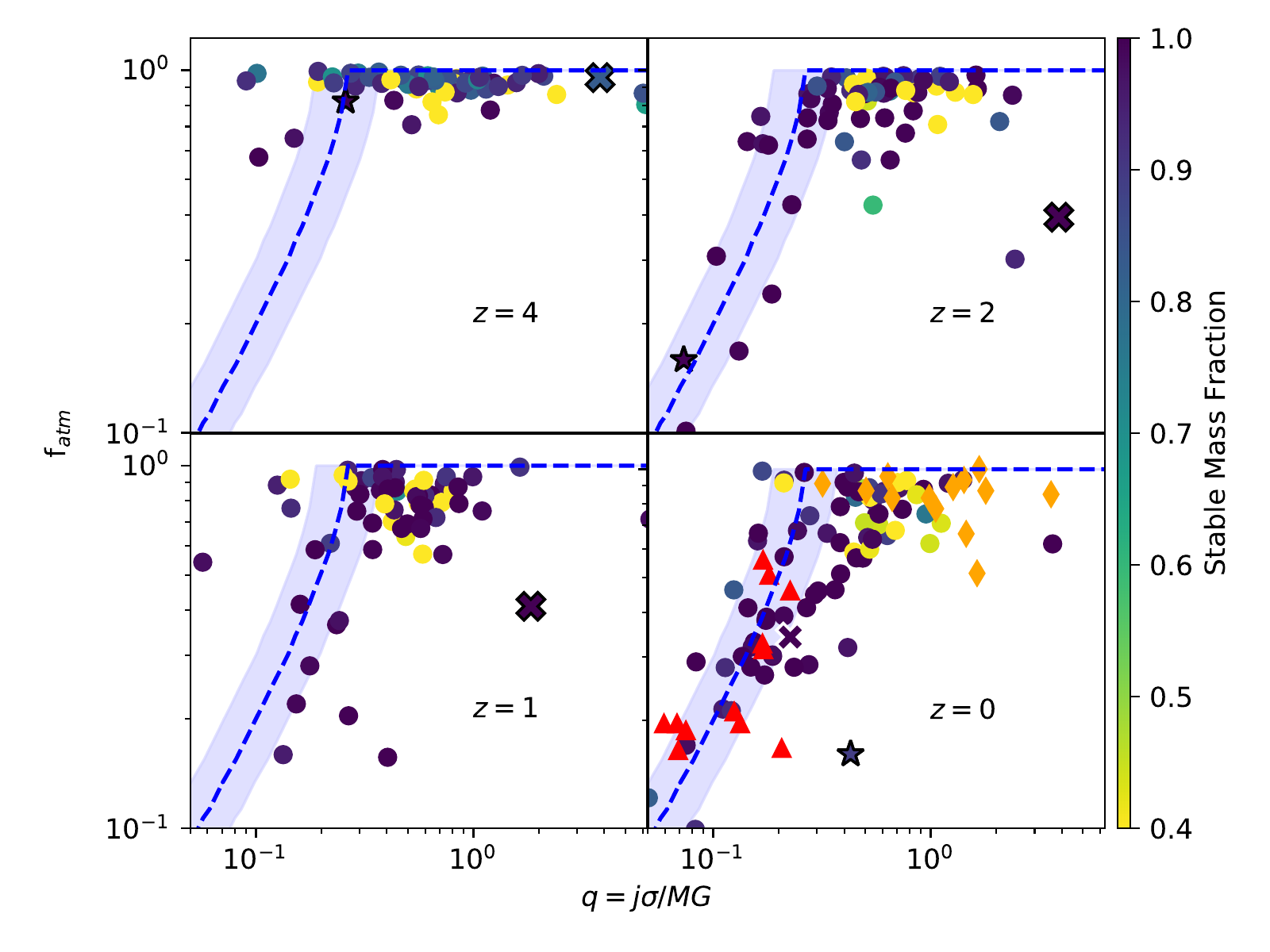}
\caption{Evolution of atomic gas fraction versus q parameter. The blue dashed line shows the O16 relation given in eq.~(\ref{eq:fatm}) with a $40\%$ uncertainty (in $f_{\rm atm}$) as the \hi dispersion exhibits an intrinsic empirical scatter \citep{obreschkow16}. All points are color-coded by the stable mass fraction of the galaxies at the given redshift. Red triangles and orange diamonds show the local spiral galaxies from the THINGS survey and the LITTLE THINGS survey analyzed by O16.  The simulated galaxies are broadly consistent with the prediction of the model of O16 and therefore confirm that the atomic gas fraction is connected to the cosmic evolution of q. Two outliers with irregular morphologies discussed in Section~\ref{ss:stats} are marked as $\star$ symbol (g8.13e11) and $\times$ symbol (g1.37e11).
         \label{fig:q_fatm_z}}
\end{figure*}

\subsection{Statistical relation between f$_{\rm atm}$ and $q$}\label{ss:stats}

The location of all 88 NIHAO galaxies in the $q$-$f_{\rm atm}$ space is shown in Fig.~\ref{fig:q_fatm_z} at four different redshifts. The dashed line shows the prediction of the equilibrium model. The simulations exhibit a redshift evolution of $f_{\rm atm}$ at fixed $q$ parameter since redshift $z \sim 4$. The  maximum deviations of simulated galaxies relative to the model of O16 are contained within $\lesssim0.5\rm~dex$, despite the six orders of magnitude in stellar mass spanned by this sample. The NIHAO simulations therefore confirm that the atomic gas fraction is connected to the cosmic evolution of $q$. There are nonetheless clear systematic deviations between the analytical model and the simulations, which we will discuss now.

Firstly, at $z=4$, all simulated galaxies exhibit very high atomic gas factions, even at the lowest $q$ values, where lower gas fractions are expected from the stability model. This is because the timescale for accretion is indeed shorter than that of the \hi-\hm transition in these few galaxies. Hence the analytical equilibrium model is bound to fail (see Section \ref{sec:overview}).

Secondly, at $q$-values larger than $q=1/(\sqrt{2}e)$, where the O16 model predicts purely atomic disks ($f_{\rm atm}=1$), the simulated galaxies fall systematically below the model. This discrepancy increases from $z=4$, where $f_{\rm atm}\approx0.9$ in this regime, to $z=0$, where $f_{\rm atm}\approx0.6$--$0.7$. Most galaxies in this range of $q$ are dwarf galaxies. In the simulation (as well as in reality) such galaxies often show irregular morphologies that defy the assumption of an axially symmetric disk and show more local instabilities than expected in such a simple model. This is one reason for the offset between the model and the simulations. However, it should be emphasized that observations of dwarf galaxies at $q>1/(\sqrt{2}e)$ at $z=0$ normally exhibit atomic gas fractions that lie indeed around $0.8 - 0.9$ \citep{obreschkow16}. It is therefore possible, that our result somewhat under-predicts the atomic gas in dwarf galaxies. In either case, the reason for the better agreement with the O16 relation at $z=4$ is that, at such high redshift, almost all galaxies have barely started to form stars.


 
Thirdly, the stable mass fraction does not correlate with the deviation between the simulated galaxy and the model of O16. Most galaxies with an unstable gaseous disk lie at $q$-values larger than $q=1/(\sqrt{2}e)$, and the fraction of such galaxies in the sample is less than $5 \%$. By inspecting the evolution of these unstable systems snapshot-by-snapshot, we found that all the most unstable systems ($f_{\rm stable}<0.5$) are only unstable for one snapshot. In other words, the timescale of the instability is shorter than the temporal resolution of the NIHAO simulation. We are therefore currently unable to determine how long exactly the unstable phases lasts.
 
Finally, two interesting outliers are marked as different symbols in Fig.~\ref{fig:q_fatm_z}.
A one-by-one inspection of these galaxies shows that they have either irregular morphologies. The NIHAO object g8.13e11 ($\star$ symbol) at redshift $z = 0$ has a polar ring at $z=0$. Object g1.37e11 ($\times$ symbol) shows a clear signature of a recent merger event at $z = 2$ and acquires a spherical morphology with faint streams at redshift $z = 1$.  


\section{Conclusions} \label{sec:summary}

In this paper, we used the NIHAO galaxy simulation suite \citep{wang15} to analyze the dependency between the atomic gas fraction $f_{\rm atm}$ and the integrated atomic stability parameter $q$ \citep{obreschkow16} across cosmic time. The $q$ parameter was defined by O16 and used to develop an analytical equilibrium model to predict the atomic gas fraction in disks. NIHAO is a large set of high resolution cosmological zoom-in
hydrodynamical galaxy formation simulations in the mass range between dwarf galaxies to Milky-Way mass galaxies. The simulated galaxies have a realistic cosmological environment and realistic dynamical and kinematic properties, making them ideal to test the O16 model in a full cosmological set-up. Our results are:

\begin{itemize}
\item The atomic gas fractions for all galaxies start at unity and decrease monotonically as the galaxies evolve. The galaxies in the most massive mass bin consume their gas rapidly while galaxies in lower mass bins decrease more mildly.
\item Most ($\gtrsim90\%$) atomic gas of most galaxies is stable at any cosmic time. Most of the stable gas is clearly stable (Toomre $Q > 2$).
\item The NIHAO sample is qualitatively consistent with the model of O16, which predicts the atomic gas fraction to depend on mass and angular momentum only via the \textit{integrated atomic stability parameter} $q$. The simulation and model agree at almost any time.
\end{itemize}

The last point is the most important finding. It implies that gravitational equilibrium is the dominant factor regulating $f_{\rm atm}$ at any particular time. The deeper reason for this simple conclusion is that the timescale of \hi accretion is almost always longer than that of the local \hi$\leftrightarrow$~\hm feedback loop. An exception to this rule are galaxies undergoing strong interactions, which can lead to massive instantaneous accretion and/or remove large amounts of \hi, for instance via starbursts, dynamical heating, stripping or fuelling of a central black hole. Some of these additional processes have recently been explored by \citet{stevens18} in a semi-analytic context, but a full physics treatment of these processes remains yet to be presented.


\acknowledgments

We thank Joop Schaye and Alessandro Romeo for useful feedback on this manuscript. We also thank the anonymous referee for a constructive report that helped improve the clarity of this paper. 
The analysis was performed using the pynbody package (http://pynbody.github.io),
which was written by Andrew Pontzen and Rok Ro{\v s}kar in addition to the authors.
CL has received funding from a Discovery Early Career Researcher Award (DE150100618) and by the ARC Centre of
Excellence for All Sky Astrophysics in 3 Dimensions (ASTRO 3D), through project number CE170100013.
This research was carried out on the High Performance Computing resources at New York University Abu Dhabi;
on the {\sc theo}  cluster of the Max-Planck-Institut f\"ur Astronomie and on the {\sc hydra}  clusters at the Rechenzentrum in Garching.

%

\bibliographystyle{apj}
\bibliography{nihao_fatmq}




\appendix

\section{Calculation of stability parameter of galaxies in NIHAO} \label{app:stab}

In order to measure the stability parameter that was shown in Section~\ref{sec:sim}, we sample the simulated galaxy with 400 cells across a plane perpendicular to its spin. In each cell, we measure the surface density $\Sigma$, local radial velocity dispersion $\sigma$ and local epicyclic frequency $\kappa$.

Gas and stellar mass surface densities were calculated from the enclosed mass within the cell as $\Sigma_{\rm gas} = M_{\rm gas} / S$ and $\Sigma_{\rm star} = M_{\rm star} / S$ where $M_{\rm gas}$ and $M_{\rm star}$ are total gaseous and stellar mass within each cell and $S$ is the area of each cell.
 
We calculate the star-forming gas velocity dispersion of galaxies by considering the velocity difference with the centre of mass, and calculating the component of this velocity that is parallel to the rotation axis:
\begin{equation}
\sigma_{\rm gas} = \sqrt[]{\frac{\sum_{\rm i} m_{\rm i} \left(v_{\rm z, i}^2 + \sigma_{\rm P}^2 \right)}{\sum m_{\rm i}}}
\end{equation}
Here, $i$ are all gas particles within each cell, $m_{\rm i}$ is the mass of particle $i$ and $v_{\rm z,i}$ is the vertical velocity of the $i$ particle with respect to the centre of mass. The velocity dispersion contribution from the gas pressure of gas particles is $\sigma_{\rm P}$ and is defined as:
\begin{equation}
\sigma_{\rm P} = \sqrt[]{\frac{P}{\rho}}
\end{equation}
where $P$ and $\rho$ are the gas pressure and density.
In the case of stars, we calculate the velocity dispersion in a similar manner, but in the case of stars there is no thermal pressure, so the stellar velocity dispersion is simply
\begin{equation}
\sigma_{\rm star} = \sqrt[]{\frac{\sum_{\rm i} m_{\rm i} v_{\rm z, i}^2}{\sum m_{\rm i}}}
\end{equation}

As simulations have full kinematic information of particles, we can measure the epicyclic frequency by definition
\begin{equation}
\kappa = \Omega + \omega
\end{equation}
where $\Omega$ is the $z$-component of angular velocity of the cell relative to the center of galaxy, and $\omega$ is the $z$-component of angular velocity of all particles within the cell relative to their mass center. The angular velocity can be calculated by $\omega = J_{\rm z} / I_{\rm zz}$ where $J_{\rm z}$ is the angular momentum of all particles within the cell and $I_{\rm zz}$ is the $zz$-component of the inertia tensor
\begin{equation}
I_{zz} = \sum^N_{k=1} m_k \left( x^2_k + y^2_k\right).
\end{equation}

To measure the \citet{toomre64} parameter for each cell, we use the properties above as:
\begin{equation}
Q_{\rm gas} = \frac{\kappa \sigma_{\rm gas}}{\pi G \Sigma_{\rm gas}},
\end{equation}
\begin{equation}
Q_{\rm star} = \frac{\kappa \sigma_{\rm star}}{\pi G \Sigma_{\rm star}}.
\end{equation}
We combine $Q_{\rm gas}$ and $Q_{\rm star}$ to get a net Toomre parameter following \citet{romeo11},
\begin{equation}
\frac{1}{Q_{\rm net}} = \begin{cases}
              \frac{W}{Q_{\rm star}} + \frac{1}{Q_{\rm gas}}, Q_{\rm star} \geq Q_{\rm gas} \\
              \frac{W}{Q_{\rm gas}} + \frac{1}{Q_{\rm star}}, Q_{\rm gas} \geq Q_{\rm star},
              \end{cases}
\end{equation}
where
\begin{equation}
W = \frac{2\sigma_{\rm gas}\sigma_{\rm star}}{\sigma^2_{\rm gas} + \sigma^2_{\rm star}}.
\end{equation}

\end{document}